\documentclass[a4paper]{article}

\usepackage{amsmath} \usepackage[psamsfonts]{amssymb}
\usepackage{cmmib57} \usepackage{amsthm}

\usepackage[dvips]{graphicx}

\def\idty{{\rm1\mkern-5.4mu I}}
\renewcommand{\Bbb}[1]{\if1#1\idty\else\mathbb{#1}\fi}
\newcommand{\scr}[1]{\mathcal{#1}}

\def\order#1{{\bf o}\left(#1\right)}

\newcommand{\kbpp}{|\psi\rangle\langle\psi|}
\newcommand{\tr}{\operatorname{tr}}
\newcommand{\U}{\operatorname{U}}
\newcommand{\SU}{\operatorname{SU}}
\newcommand{\Sym}{\operatorname{S}}
\newcommand{\fidopt}{\scr{F}^{\rm max}}
\newcommand{\rate}{\mu}
\newcommand{\noise}{\lambda}
\newcommand{\natpur}{T^{\rm nat}}
\newcommand{\optpur}{T^{\rm opt}}

\newtheorem{thm}{Theorem}[section]
\newtheorem{lem}{Lemma}[section]
\newtheorem{prop}{Proposition}[section]
\newtheorem{defi}{Definition}[section]

\title{The Rate of Optimal Purification procedures}
\author{M. Keyl\thanks{Electronic Mail: \texttt{m.keyl@tu-bs.de}} and
  R.~F. Werner\thanks{Electronic Mail: \texttt{R.Werner@tu-bs.de}}
  \\[1ex]
  {\small Institut f{\"u}r Mathematische Physik, TU Braunschweig,}\\
  {\small Mendelssohnstr.3, 38106 Braunschweig, Germany.}}
\date{\today}

\begin{document}
\maketitle

\begin{abstract}
 Purification is a process in which decoherence is partially
reversed by using several input systems which have been subject to
the same noise. The purity of the outputs generally increases with
the number of input systems, and decreases with the number of
required output systems. We construct the optimal quantum
operations for this task, and discuss their asymptotic behaviour
as the number of inputs goes to infinity. The rate at which output
systems may be generated depends crucially on the type of purity
requirement. If one tests the purity of the outputs systems one at
a time, the rate is infinite: this fidelity may be made to
approach 1, while at the same time the number of outputs goes to
infinity arbitrarily fast. On the other hand, if one also requires
the correlations between outputs to decrease, the rate is zero: if
fidelity with the pure product state is to go to 1, the number of
outputs per input goes to zero. However, if only a fidelity close
to 1 is required, the optimal purifier achieves a positive rate,
which we compute.

\end{abstract}

\section{Introduction}
\label{sec:intro}

A central problem of quantum information processing is to ensure
that devices which have been designed to perform certain tasks
still work well in the presence of decoherence, i.e., under the
combined influences of inaccurate specifications, interaction with
further degrees of freedom, and thermal noise. Decoherence
typically has the effect of producing mixed states out of pure
states, so it is natural to ask whether the effects of decoherence
can be partially undone, by processes turning mixed states into
purer ones. As in the classical case this is impossible for
operations working on single systems. However, if many (say $N$)
systems are available, all of which were originally prepared in
the same unknown pure state $\sigma$, and subsequently exposed to
the same (known) decohering process $R_*$, then an analysis of the
combined state may well allow the reconstruction of the original
pure state. The quality of this reconstruction will increase with
$N$. In fact, it should approach perfection as $N\to\infty$: in
this limit one can determine the decohered state $R_*\sigma$ to an
arbitrary accuracy by statistical measurements. The question is
only, whether the knowledge of the full density matrix $R_*\sigma$
admits the reconstruction of $\sigma$, i.e., whether the linear
operator $R_*$ is invertible. Generically, and for sufficiently
small decoherence, this is the case. However, the operator
$R^{-1}_*$ is usually not positive, i.e., it takes some density
matrices into operators with negative eigenvalues. Therefore, it
does not correspond to a physically realizable apparatus. But it
does describe a computation we can perform to reconstruct $\sigma$
from the measured (or estimated) density matrix $\rho=R_*\sigma$.

How well can this reversal of decoherence be done when the number
$N$ of inputs is given, and finite? The answer depends critically
on the way the purification task is set up, and what ``figure of
merit'' we try to optimize. In general, the resulting variational
problems may be very hard to solve. However, in the specific model
situation chosen in this paper, the solution is fairly
straightforward: we take qubit systems, and assume that
decoherence is described by a depolarizing channel of the form
\begin{equation}
  \label{eq:1}
  R_*\sigma= \lambda\sigma+(1-\lambda){\idty\over2}.
\end{equation}
The purifier will be a device $T$ taking a state of $N$ qubits,
and turning out some number $M$ of qubits, where $M$ may be either
fixed or itself a random quantity. In the latter case $T$ is given
mathematically by a family $T_M$ of completely positive maps,
where $T_M$ takes a density matrix of $N$ qubit systems, and
produces a positive operator on the $M$ qubit space, which is not
necessarily normalized to unity: the normalization constant
$w_M=\tr(T_M(\rho))$ is interpreted as the probability of getting
exactly $M$ outputs from the input state $\rho$. Thus
$\sum_Mw_M=1$.

Our aim is to design $T$ to get outputs as close as possible to
the uncorrupted input state $\sigma$, and also as many of them as
possible. This is reminiscent of cloning problems
\cite{Klo1,Klo2}. However, in cloning problems the aim is to get
many copies of the input state to $T$, which in our case is the
mixed state $R_*\sigma$, rather than the pure state $\sigma$. In
both cases there is clearly a trade-off between the quality
of the outputs and their number, which is why there are several
different ways to state the problem. In the sequel we will briefly
describe the variants of the purification problem, together with
the results, which will be shown later in the paper.

\begin{enumerate}
\item \label{it:0}
  \textit{Maximal fidelity, failure to produce any output admissible.}
  The best fidelity of outputs is clearly achieved, when the weakest
  possible demands are made on the number of outputs. In this case we
  do not even insist on an output every time the device is run, but
  only on some non-zero probability for getting an output. The
  best achievable fidelity of these outputs goes to $1$ as
  $N\to\infty$, but not substantially faster than with the
  following stronger requirement on output numbers.

\item \label{it:1}
   \textit{$M=1$ fixed, number $M$ never increased at expense
    of output purity.}
  This is the approach taken by \cite{IAC}. At least one output qubit is
  required, and the figure of merit is based on the fidelity of
  this one qubit. As it turns out the optimal device for this
  problem can just as well produce more outputs of the same
  optimal fidelity, with a certain rate. However, this rate
  is not part of the optimization criterion.

\item \label{it:2}
  \textit{$M$ fixed, purity measured by one-particle
    restrictions.} For fixed $M,N$, this problem is rather similar to
  \ref{it:1}. However, with the additional parameter $M$ we can
  discuss better the trade-off between rate and quality of
  outputs. Suppose we fix some dependence of the number of outputs
  $M(N)$ on the number of inputs. Do the states still approach $\sigma$
  as $N\to\infty$? Clearly, if $M(N)$ increases
  slowly, e.g., at the rate given by the optimal device from
  \ref{it:1}, this will be the case. What may seem surprising at
  first, however, is that {\it no matter} how fast $M(N)\to\infty$,
  the state of each output qubit still approaches the uncorrupted pure
  state. In this sense, optimal purification works with an {\it infinite
  rate}.

\item\label{it:3}
\textit{$M$ fixed, purity measured by fidelity with respect to
  $\sigma^{\otimes N}$.}
  The infinite rate depends critically on what we use as the quality
criterion for outputs. Apart from the fidelity of the restrictions
of the output state to single qubits used in \ref{it:2} we could
also look at the fidelity of the outputs with respect to the $M$
particle pure state $\sigma^{\otimes M}$, thereby taking into
account also the correlations between different outputs. For fixed
$M$, the difference between these two fidelity measures does not
seem so great, because one can be estimated in terms of the other.
However, the estimates are $M$-dependent (see below), and hence
for problems involving a limit $M\to\infty$ the fidelity with
respect to the combined state may (and does) turn out to be a much
tighter criterion. In fact, no process with finite rate $M/N$
achieves fidelity$\to1$, and in this sense even optimal
purification works with {\it zero rate}, in sharp contrast to
\ref{it:2} above. On the other hand, for any finite fidelity
requirement, there is an output rate for an optimized process,
which is computed below.

\end{enumerate}

\noindent These results will be stated in precise terms in the
following Section~\ref{sec:merit}, together with the notation
needed for that purpose, and graphs of the optimal fidelities and
rates. The proofs follow in the subsequent sections. Technically
they hinge on the decomposition theory of tensor product
representations of SU(2), and this background is provided in
Section~\ref{sec:decomp}. The reason for representation theory to
enter in such a crucial way is isolated in
Section~\ref{sec:decomp-reduc}, where it is shown that the optimal
devices can be taken to be SU(2)-covariant (do not single out a
basis in the qubit space). The two basic purifiers, called the
``natural purifier'' (optimal for question \ref{it:1} above), and
the ``optimal purifier'' (optimal for question \ref{it:2} above)
are defined in Section~\ref{sec:Natur-optim-purif}, and their
fidelities are computed. The proof of the optimality claims is
given in Section~\ref{sec:Solut-optim-probl}. Finally, in
Section~\ref{sec:asympt}, we determine the asymptotic behaviour
for the optimal purifier, and the output rates.

\section{Figures of Merit and Main Results}
\label{sec:merit}

In this section we will state the optimization problems for
purifiers mathematically. A device (not necessarily a purification
procedure) taking $N$ qubit systems as input and producing $M$
output qubits is described mathematically by a trace preserving,
completely positive linear map (``cp-map'')
\begin{displaymath}
  T_*: \scr{B}_*(\scr{H}^{\otimes N}) \to \scr{B}_*(\scr{H}^{\otimes M}),
\end{displaymath}
which takes input density matrices to output density matrices.
Equivalently, we may work in the Heisenberg picture, using the
dual $T$ of $T_*$, the \emph{unital} (i.e. $T(\Bbb{1}) = \Bbb{1}$)
cp-map
\begin{displaymath}
   T: \scr{B}(\scr{H}^{\otimes M}) \to \scr{B}(\scr{H}^{\otimes N}),
\end{displaymath}
which is related to $T_*$ by $\tr\bigl(T(X)\rho\bigr) =
\tr\bigl(XT_*(\rho)\bigr)$. Here $\scr{H} = \Bbb{C}^2$ is the one
qubit Hilbert space, $\scr{B}(\,\cdot\,)$ is
the space of all (bounded) operators on the corresponding Hilbert
space and $\scr{B}_*(\,\cdot\,)$
 denotes the space of trace class operators.
Since $\dim\scr{H}=2<\infty$, the spaces $\scr{B}_*(\scr{H})$ and
$\scr{B}(\scr{H})$ are just the $2\times2$-matrices, but it is
nevertheless helpful to keep track of the distinction between
spaces of observables and spaces of states.

``Good purifiers'' should make $T_*((R_*\sigma)^{\otimes N})$ very
close to $\sigma^{\otimes M}$. A simple figure of merit is the
fidelity of the output with respect to the desired state in the
worst case, i.e.,
\begin{equation}\label{Fall}
   {\scr F}_{\rm all}(T)=\inf_\sigma
      \tr\left(\sigma^{\otimes M}T_*\bigl((R_*\sigma)^{\otimes N}
           )\bigr)\right),
\end{equation}
 where the infimum is over all one-particle pure states $\sigma$.
Similarly, we could pick any one of the outputs, say the one with
number $i$, $1\leq i\leq M$, and test its fidelity. The worst case
then gives the fidelity
\begin{equation}\label{Fone}
   {\scr F}_{\rm one}(T)=\inf_i\ \inf_\sigma
      \tr\left(\sigma^{(i)}T_*\bigl((R_*\sigma)^{\otimes N})
            \bigr)\right),
\end{equation}
where
$\sigma^{(i)}=\idty\otimes\cdots\otimes\sigma\otimes\cdots\otimes\idty$
denotes the tensor products with $(M-1)$ factors ``$\idty$'' and
one factor $\sigma$ at the $i^{\rm th}$ position. We seek to
maximize these numbers by judicious choice of $T$. Let us denote
the optimal values by
\begin{equation} \label{Fopt}
   \fidopt_\sharp(N,M)= \sup_T \ {\scr F}_{\sharp}(T),
\end{equation}
where $\sharp$=``all'' or $\sharp$=``one'', and the supremum is
over all unital cp-maps $T$ with the specified number of inputs
and outputs.

For devices with \textit{variable numbers of outputs} all these
quantities become random variables, as well. Typically, one will
seek to optimize the mean fidelity. It is then natural not to take
the infimum in Equation (\ref{Fone}), but the mean. The case where no
output is produced at all, is interpreted here as one output qubit in
the completely mixed state. The resulting \textit{mean fidelity}
\cite{IAC} can be thought of as the fidelity ${\scr F}_{\rm
  one}(\widetilde T)$ of a modified device $\widetilde T$, which uses
$T$, followed by a random selection of one of the outputs. Therefore,
the problem of maximizing mean fidelity is exactly the same as
maximizing ${\scr F}_{\rm one}(T)$ for devices with fixed output
number $M=1$, with optimal value $\fidopt_\sharp(N,M)$.

Rather than looking at the mean of the fidelity distribution of a
device with variable number of outputs we could also look at its
maximum. This corresponds to the problem in item~\ref{it:0} of the
previous section. More precisely, one should omit the ``worst
case'' infimum with respect to $i$ in this case, and allow the
device to either pick one of its outputs, or to declare failure.
This leads to a device with only the two output numbers $0$ and
$1$, and the functional to be optimized is the fidelity of the
``$1$''-output. We will denote the optimum for this problem by
$\fidopt_\sharp(N,0)$, with a slight abuse of notation expressing
that this is the case with no demands on output numbers at all.

It is clear that $\fidopt_\sharp(N,M)$ is a decreasing function of
$M$, and that therefore the limit
\begin{equation}
   \fidopt_{\sharp}(N,\infty)= \lim_{M\to\infty}\ \fidopt_{\sharp}(N,M)
\end{equation}
exists. For $\sharp$=all, this limit is zero. However, for
$\sharp$=one, it is an interesting quantity, which even goes to $1$
as $N\to\infty$.

\begin{figure}[htbp]
  \begin{center}
    \includegraphics[scale=1]{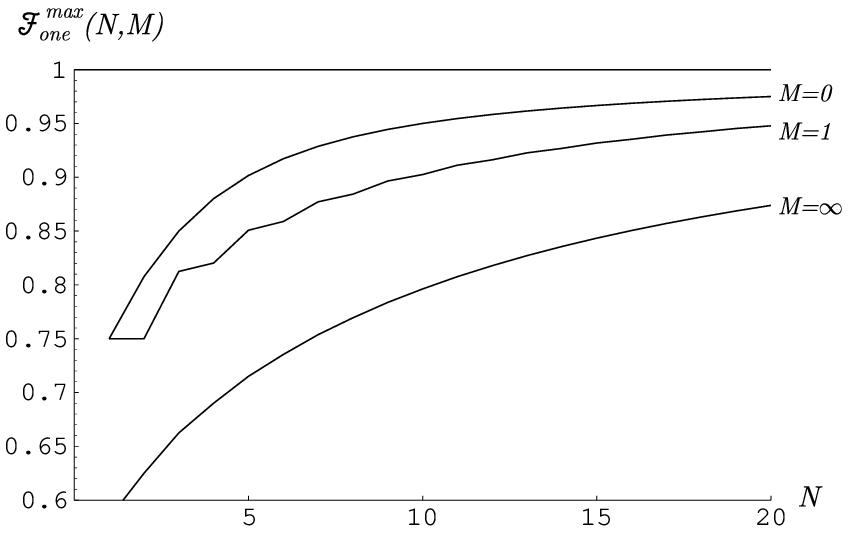}
    \caption{The three basic fidelities for the one-particle figure of
    merit:\newline
    top: $\fidopt_{\rm one}(N,0)$, middle: $\fidopt_{\rm one}(N,1)$,
    bottom: $\fidopt_{\rm one}(N,\infty)$}
    \label{fig:one}
  \end{center}
\end{figure}

The results for the quantities $\fidopt_{\rm one}(N,0)$,
$\fidopt_{\rm one}(N,1)$, and $\fidopt_{\rm one}(N,\infty)$ are
shown in  Figure~\ref{fig:one}. Of course, all these  quantities
also depend on the parameter describing the noise, which we have
suppressed for notational convenience. It is fixed in the
following graphs as $\noise=0.5$ (resp. $\beta=0.549$, see
Section~\ref{sec:decomp}). It is clear that
$\fidopt_\sharp(N,M)\to1$ for any $N$ and $M$, as the noise level
goes to zero ($\noise\to1$).

The leading asymptotic behaviour (as $N\to\infty$) is of the form
\begin{eqnarray}
   \fidopt_{\rm one}(N,M)
           &\propto& 1- \frac{c_M}{2N} + \cdots \label{eq:cs}\\
    c_0&=&(1-\lambda)/\lambda  \label{eq:c0}\\
    c_1&=&(1-\lambda)/\lambda^2\label{eq:c1}\\
    c_\infty&=&(\lambda+1)/\lambda^2.\label{eq:ci}
\end{eqnarray}

\begin{figure}[htbp]
  \begin{center}
    \includegraphics[scale=1]{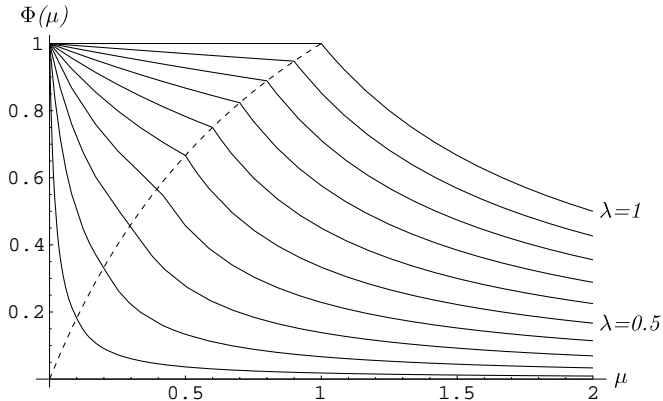}
    \caption{Asymptotic fidelity $\Phi(\mu)$ for the all-particle figure of
    merit (\ref{eq:20}).\newline
    Curve parameter:  $\noise=.1,.2,...,1$; dotted line: natural
    purifier}
    \label{fig:alltest}
  \end{center}
\end{figure}

From these asymptotic results, a simple estimate for the
all-particle fidelity criteria can be obtained: By Equation
(\ref{Fone<>Fall}),
 $1-{\scr F}_{\rm all}(T) \leq M(1-{\scr F}_{\rm one}(T))$,
where $M$ is the number of outputs. Hence, for sufficiently small
rate $M/N$ one achieves good fidelity, even for the all-particle
test criterion:
 $ 1-\fidopt_{\rm all}(N,M)
     \leq M(1-\fidopt_{\rm one}(N,M))
     \leq M(1-\fidopt_{\rm one}(N,\infty))
     \approx\frac{M}{2N}c_\infty$.
Of course, the second estimate is rather crude, and a refined
version will be given in Section~\ref{sec:asympt}. The argument
does show, however, that one may expect optimal all-particle
fidelity to become a function of the output rate. This function
will be computed in Section~\ref{sec:asympt-all}): for  every
$\rate>0$, we find the limit
 \begin{equation} \label{eq:20}
  \Phi(\rate)=\lim_{N\to\infty\atop M/N\to\rate}\
                  \fidopt_{\rm all}(N,M)
     = \begin{cases}
            \displaystyle\frac{2\noise^2}{2\noise^2 + \rate(1-\noise)} &
                            \mbox{if $\rate \leq \noise$}\\
            \displaystyle\frac{2\noise^2}{\rate(1+\noise)} &
                           \mbox{if $\rate \geq \noise$.}\\
       \end{cases}
\end{equation}
The function $\Phi$ is continuous and satisfies $\Phi(0)=1$ and
$\Phi(\infty)=0$, so at small rates purification is near perfect,
but becomes arbitrarily bad at too high rates. In
Figure~\ref{fig:alltest} $\Phi$ is plotted with the noise
parameter $\noise$ going in steps of $0.1$ from  $0$ to $1$. The
dotted line describes the performance of the natural purifier (see
Section~\ref{sec:natpur}), which operates with rate
$\rate=\noise$.

\section{Decomposition theory} \label{sec:decomp}

Many arguments in this paper are based on group theory, in
particular the decomposition of tensor products of irreducible
representations of $\SU(2)$. In this section we will summarize the
relevant results which are needed throughout the paper.

\subsection{Reduction to fully symmetric case}
\label{sec:decomp-reduc}

There are two reason why group theory is useful for us. First
of all the depolarizing channel $R$ producing the noise is
``covariant'' which means that it does not prefer any particular
polarization direction (basis in the underlying Hilbert space ${\cal
  H}=\Bbb{C}^2$), and second we are looking at a ``universal''
purification problem, i.e. the purification devices $T$ we are looking
for should work well on an arbitrary unknown input state $\sigma$.
Therefore, it is natural to look at those $T$ which are covariant as
well: $T$ should work in exactly the same way on any input.
Carrying this idea further it should also be impossible to single out
any one of the input and output channels. Mathematically, these
``natural conditions'' are stated as follows:

\begin{defi} \label{def:full-sym}
  A unital, cp-map
$T: \scr{B}(\scr{H}^{\otimes M}) \to \scr{B}(\scr{H}^{\otimes N})$
is called \emph{fully symmetric} if it is $\U(2)$ covariant, i.e.
\begin{displaymath}
  T(U^{\otimes M} A U^{*\otimes M}) = U^{\otimes N}T(A)U^{*\otimes N}
\quad \forall A \in%
  \scr{B}(\scr{H}^{\otimes M}) \ \forall U \in \U(2)
\end{displaymath} and permutation invariant, i.e.
\begin{displaymath}
  T(\eta A \eta^*) = T(A) \quad \forall \eta \in \Sym_M \
  \forall A \in \scr{B}(\scr{H}^{\otimes M})
\end{displaymath} and
\begin{displaymath}
  \tau T(A) \tau^* = T(A) \quad \forall \tau \in \Sym_N \ %
\forall A \in \scr{B}(\scr{H}^{\otimes N}).
\end{displaymath}
 Here $\eta \in \Sym_M$, $\tau \in \Sym_N$ denote permutations of $M$
respectively $N$ elements and at the same time the corresponding
unitaries on $\scr{B}(\scr{H}^{\otimes M})$ and
$\scr{B}(\scr{H}^{\otimes N})$, i.e. $\eta (\psi_1 \otimes \cdots
\otimes \psi_M) = \psi_{\eta(1)} \otimes \cdots \otimes
\psi_{\eta(M)}$.
\end{defi}

We could have made this condition part of our definition of a
purifier, and restricted the discussion to fully symmetric
operations from the outset. However, we have chosen to take the
heuristic arguments at the beginning of this section more
seriously: the kind of ``universality'' described there is already
embodied in the figures of merit of Section~\ref{sec:merit}, so it
becomes a mathematical question whether optimal purifiers are
indeed fully symmetric or else symmetry is broken, and a
non-symmetric purifier can outperform all symmetric ones.

We now argue that the \emph{optimal} devices (with respect  to
$\scr{F}_{\rm one}$ and $\scr{F}_{\rm all}$) may be indeed by
assumed to be fully symmetric. To make this precise, note that
$\scr{F}_{\rm all}(T)$ and $\scr{F}_{\rm one}(T)$ are infima over
expressions which are linear in $T$, and hence concave
functionals. Therefore, averaging over many $T$'s with the same
figure of merit produces a $T$ at least as good. Clearly, for all
permutations  $\eta \in \Sym_M$, $\tau \in \Sym_N$ and
$U\in\U(2)$, the purifier $T'(X)=\tau U^{\otimes N}T(\eta\, U^{*
\otimes M}X U^{\otimes M}\eta^*)U^{* \otimes N}\tau^*$ has the
same figure of merit as $T$. By averaging over these parameters
(with respect to the appropriate Haar measures) we thus find a
purifier, which is at least as good as $T$ and, in addition,
fully symmetric. Similar arguments apply for purifiers with variable
numbers of outputs (although one has to be more careful in
defining figures of merit). Therefore, we will restrict our
discussion to fully symmetric purifiers from now on.

\subsection{Decomposition of tensor products}
\label{sec:decom-oper}

The reduction to fully symmetric purifiers allows the application
of techniques from group theory (especially representation theory
of $\SU(2)$) which simplifies our problems significantly. Consider
in particular the $N-$fold tensor product
\begin{displaymath}
  \SU(2) \ni U \mapsto \pi_{1/2}(U)^{\otimes N}=U^{\otimes N}
      \in \scr{B}(\scr{H}^{\otimes N}),
\end{displaymath} of the spin-$1/2$, or the ``defining'' representation
$\SU(2) \ni U \mapsto \pi_{1/2}(U) = U \in \scr{B}(\scr{H})$. It
decomposes into a direct sum of irreducible subrepresentations
\begin{equation}\label{eq:pidecomp}
  \pi_{1/2}(U)^{\otimes N} =U^{\otimes N}
     = \bigoplus_{s \in I[N]} \pi_s(U) \otimes \Bbb{1}
\end{equation} with
\begin{displaymath}
  \pi_s(U) \otimes \Bbb{1} \in \scr{B}(\scr{H}_s \otimes \scr{K}_{N,s}) \
\text{and}
  \ \scr{H}^{\otimes N} = \bigoplus_{s\in I[N]} \scr{H}_s \otimes
\scr{K}_{N,s}
\end{displaymath} and
\begin{displaymath}
  I[N] =
  \begin{cases}
    \{ 0,1,\ldots,\frac{N}{2} \} & \mbox{$N$ even} \\
    \{ \frac{1}{2}, \frac{3}{2} \ldots, \frac{N}{2} \} & \mbox{$N$ odd}
  \end{cases}
\end{displaymath} Here $\pi_s$ denotes the spin-$s$ irreducible
representation of $\SU(2)$, $\scr{H}_s$ its $2s+1$-dimensional
representation space, which we will identify in the following with
the symmetric tensor-product $\scr{H}_+^{\otimes2s}$, i.e. the
$2s$--qubits Bose subspace, and $\scr{K}_{N,s}$ denotes a
multiplicity space, which carries an appropriate representation of
the symmetric group $\Sym_N$.

\subsection{Decomposition of states} \label{sec:decom-st}

Consider now a general qubit density matrix $\rho$, which in its
eigenbasis can be written as $(\beta \geq 0)$
\begin{align}
  \rho(\beta) &=\frac{1}{2\cosh(\beta)} \exp\left(2\beta
\frac{\sigma_3}{2}\right) =
  \frac{1}{e^\beta+e^{-\beta}}
\left(\begin{array}{ll}e^\beta&0\\0&e^{-\beta}
  \end{array}\right) \label{eq:14}  \\
  &=\tanh(\beta)\kbpp +(1-\tanh(\beta)){1\over2}\idty, \quad \psi =
  \left(\begin{array}{l} 1 \\ 0 \end{array}\right) \notag
\end{align} The parametrization of $\rho$ in terms of the
``pseudo-temperature'' $\beta$ is chosen here, because it is, as
we will see soon, very useful for calculations. The relation to
the form of $\rho = R_*\sigma$ initially given in Equation (\ref{eq:1}) is
obviously \begin{displaymath}
  \noise = \tanh(\beta).
\end{displaymath}
The $N$--fold tensor product $\rho^{\otimes N}$ can be expressed
as $\rho(\beta)^{\otimes N} = (2 \cosh(\beta))^{-N} \exp(2\beta L_3)$ where
\begin{equation} \label{eq:3}
  \scr{B}(\scr{H}^{\otimes N}) \ni L_3
     =  \frac{1}{2} \left( \sigma_3 \otimes
    \Bbb{1}^{\otimes(N-1)} + \cdots + \Bbb{1}^{\otimes(N-1)} \otimes
\sigma_3 \right).
\end{equation}
 denotes the 3--component of angular momentum in the
representation $\pi_{1/2}^{\otimes N}$. In other words, the
density matrices are just analytic continuations of group
unitaries, or ``SU(2)-rotations by an imaginary angle $2i\beta$''.
This reduces the decomposition of $\rho(\beta)^{\otimes N}$ to the
decomposition (\ref{eq:pidecomp}) of the tensor product
representation. Of course, analytically continued group elements
are not normalized as density operators. Extracting appropriate
normalization factors the decomposition becomes
\begin{displaymath}
  \rho(\beta)^{\otimes N}
  = \bigoplus_{s \in I[N]} w_N(s) \rho_s(\beta) \otimes
           \frac{\Bbb{1}}{\dim \scr{K}_{N,s}},
\end{displaymath} with
\begin{equation}\label{eq:2}
   w_N(s) =
\frac{\sinh\bigl((2s+1)\beta\bigr)}{\sinh(\beta)(2\cosh(\beta))^N}
       \dim\scr{K}_{N,s},
\end{equation} and
\begin{displaymath}
   \rho_s(\beta) = \frac{\sinh(\beta)}{\sinh\bigl((2s+1)\beta\bigr)}
\exp(2\beta L_3^{(s)}).
\end{displaymath} Here $L_3^{(s)}$ denotes again the 3--component
of angular momentum, now in the representation $\pi_s$.

The $\rho_s(\beta)$ are normalized, i.e. $\tr\rho_s(\beta) =1$.
Hence $\sum_s w_N(s) = 1$ and $0\leq w_N(s) \leq 1$ due to the
normalization of $\rho(\beta)^{\otimes
  N}$. Together with the fact that the multiplicities $\dim
\scr{K}_{N,s}$ are independent of $\beta$ we can extract from
Equation (\ref{eq:2}) a generating functional for
$\dim\scr{K}_{N,s}$:
\begin{align*}
     2\sinh(\beta)(2\cosh(\beta))^N
     &=2\sum_{s \in I[N]}\sinh\bigl((2s+1)\beta\bigr)\dim\scr{K}_{N,s}\\
   =\bigl(e^\beta-e^{-\beta}\bigr)\bigl(e^\beta+e^{-\beta}\bigr)^N
   &=\sum_{s \in I[N]}\left(e^{(2s+1)\beta}-e^{-(2s+1)\beta}\right)
       \dim\scr{K}_{N,s},
\end{align*} obtaining \begin{displaymath}
  \dim\scr{K}_{N,s}=\frac{2s+1}{N/2+s+1}{N\choose N/2-s},
\end{displaymath} provided $N/2-s$ is integer, and zero otherwise.
The same result can be derived using representation theory of the
symmetric group; see \cite{Simon}, where the more general case
$\dim\scr{H} = d \in \Bbb{N}$ is studied.

\subsection{Decomposition of operations and optimal cloning}
\label{sec:decomp-cpmaps}

Let us come back now to fully symmetric cp-maps $T:
\scr{B}(\scr{H}^{\otimes M}) \to \scr{B}(\scr{H}^{\otimes N})$.
Using the results of Subsection \ref{sec:decom-oper} it is easy to
see that $T$ can be decomposed into a direct sum \begin{equation}
\label{eq:8}
  T(A) = \bigoplus_{s \in I[N]} T_s(A) \otimes \Bbb{1}
\end{equation} where the $T_s:\scr{B}(\scr{H}^{\otimes M}) \to
\scr{B}(\scr{H}_s)$ are unital cp-maps which are again fully
symmetric (using an obvious modification of Definition
\ref{def:full-sym}). Identifying, as in Subsection
\ref{sec:decom-oper}, the representation space $\scr{H}_s$ with
the $2s$--fold symmetric tensor product $\scr{H}_+^{\otimes 2s}$,
leads to the significantly simpler problem of decomposing fully
symmetric, unital cp-maps $Q: \scr{B}(\scr{H}^{\otimes M}) \to
\scr{B}(\scr{H}_+^{\otimes N})$, which is already solved in
\cite{Klo2}. Hence we will state only the corresponding results
here. In particular we have the following theorem:

\begin{thm} \label{thm:opt-clo}
  Consider again the 3-components of angular momentum $L_3$ and
  $L_3^{(s)}$ in the representations $\pi_{1/2}^{\otimes M}$ respectively
  $\pi_s$ (cf. Subsection \ref{sec:decom-st}).
  \begin{enumerate}
  \item
    For each fully symmetric cp-map $Q: \scr{B}(\scr{H}^{\otimes M}) \to
    \scr{B}(\scr{H}_+^{\otimes 2s})$ there is a constant $\omega(Q) \in
    \Bbb{R}^+$ with
    $Q(L_3) = \omega(Q) L_3^{(s)}$.
  \item
    For each $2s \in \Bbb{N}_0$ there is exactly one fully symmetric
    $\hat Q_{2s}$ with
    \begin{equation} \label{eq:25}
      \omega(\hat Q_{2s})  = \max_Q \omega(Q) =
      \begin{cases}
        \displaystyle \frac{M}{2s} & \mbox{for $2s \geq M$} \\[10pt]
        \displaystyle \frac{M+2}{2s+2} & \mbox{for $2s < M$},
      \end{cases}
    \end{equation}
    where the maximum is taken over the set of all fully symmetric
    cp-maps $Q: \scr{B}(\scr{H}^{\otimes M}) \to \scr{B}(\scr{H}_+^{\otimes 2s})$.
  \item \label{item:1}
    If $M >2s$ holds $\hat Q_{2s}$ is given in terms of its pre-dual
    $\hat Q_{2s*}: \scr{B}_*(\scr{H}_+^{\otimes 2s}) \to \scr{B}_*(\scr{H}^{\otimes
      M})$ by
    \begin{equation} \label{eq:24}
      \hat Q_{2s*}(\theta) = \frac{2s+1}{M+1} S_M(\theta \otimes \Bbb{1}^{\otimes (M-2s)})S_M
    \end{equation}
    where $S_M$ is the projector from $\scr{H}^{\otimes M}$ onto the Bose
    subspace $\scr{H}^{\otimes M}_+$.
  \item \label{item:2}
    For $M \leq 2s$ the map $\hat Q_{2s}$ is given by
    \begin{displaymath}
      \hat Q_{2s}(A) = S_{2s} (A \otimes \Bbb{1}^{\otimes (2s-M)}) S_{2s},
    \end{displaymath}
    or in terms of its predual
    \begin{equation} \label{eq:26}
      \hat Q_{2s*} (\theta) = \tr_{2s-M} \theta,
    \end{equation}
    where $\tr_{2s-M}$ denotes the partial trace over the first $2s-M$
    tensor factors.
  \end{enumerate}
\end{thm}

Note that the family of cp-maps $\hat Q_2s$ defined in Equation
(\ref{eq:24}) respectively  (\ref{eq:26}) plays a very special
role not only mathematically: $\hat Q_{2s}$ describes the optimal way
to \emph{increase} (\ref{eq:24}) or \emph{dercrease} (\ref{eq:26})
the number of qubits. More precisely $\hat Q_{2s*}$  maps a finite
number $2s$ of qubits in the same unknown pure state $\sigma$ to
the best possible approximation $Q_{2s*}(\sigma^{\otimes 2s})$ of
the product state $\sigma^{\otimes M}$. The quality of
$Q_{2s*}(\sigma^{\otimes 2s})$ is measured here by the fidelities
\begin{displaymath}
  \scr{G}_{\rm all}(Q) :=  \inf_\sigma \tr\left(\sigma^{\otimes
      M}Q_*\bigl(\sigma^{\otimes 2s}\bigr)\right)
\end{displaymath}
or
\begin{displaymath}
  \scr{G}_{\rm one}(Q) :=  \inf_i\ \inf_\sigma
  \tr\left(\sigma^{(i)}Q_*\bigl(\sigma^{\otimes 2s}\bigr)\right).
\end{displaymath}
If $2s\geq M$ holds (item \ref{item:2}) we simply
have to discard $2s-M$ qubits to get exactly
$\hat Q_{2s*}(\sigma^{\otimes 2s}) = \sigma^{\otimes M}$. If the number
of qubits should be increased, i.e. $M>2s$ holds (item
\ref{item:1}), the target state $\sigma^{\otimes M}$ can not be
reached. In this case $\hat Q_{2s}$ is the \emph{optimal quantum
  cloning device} described in \cite{Klo1,Klo2}.

\section{Natural and optimal purifiers}
\label{sec:Natur-optim-purif}

In this section we will introduce a particular class of
purification maps which arise very naturally from the group
theoretical discussion of the last section and which maximize, as
we will see in Section \ref{sec:Solut-optim-probl}, the fidelities
$\scr{F}_{\rm   all}$ and $\scr{F}_{\rm  one}$.

\subsection{The definitions} \label{sec:natpur}

As a first step let us reinterpret the decomposition of
$\rho(\beta)^{\otimes N}$ discussed in Subsection
\ref{sec:decom-st} in terms of the of cp-map
\begin{multline}\label{eq:28}
  \bigoplus_{s \in I[N]} \scr{B}(\scr{H}^{\otimes 2s}_+)
  \ni \bigoplus_{s \in I[N]} A_s
  =: A \mapsto\natpur(A) :=
  \bigoplus_{s \in I[N]} \natpur_s(A_s) := \\
  := \bigoplus_{s \in I[N]}
      A_s \otimes \Bbb{1} \in \bigoplus_{s \in I[N]} \scr{B}(\scr{H}_s \otimes \scr{K}_{N,s})
      = \scr{B}(\scr{H}^{\otimes N}).
\end{multline}
Its predual maps the density matrix $\rho(\beta)^{\otimes N}$ to
$\bigoplus_{s \in I[N]} w_N(s) \rho_s(\beta)$. The latter should
be interpreted as a (normal) state on the von Neumann algebra
$\bigoplus_{s \in I[N]} \scr{B}(\scr{H}^{\otimes 2s}_+)$. Hence
$\natpur$ is an \emph{instrument} which produces with probability
$w_N(s)$ the $2s$--qubit state $\rho_s(\beta)$ from the input
state $\rho(\beta)^{\otimes
  N}$. This implies in particular that the number of output systems
of $\natpur$ is not a fixed parameter but an observable. We will
see soon that the fidelities of the output states $\rho_s(\beta)$
are bigger than those of the input state $\rho(\beta)^{\otimes N}$
provided $s > 0$ holds. Hence we will call $\natpur$ the
\emph{natural purifier}.

The most obvious way to construct a device which produces always
the same number of output systems is the composition of $\natpur$
with the cloning operation
\begin{displaymath}
  \scr{B}(\scr{H}^{\otimes M}) \ni A \mapsto \hat Q(A)
  = \bigoplus_{s \in I[N]} \hat Q_{2s}(A)
  \in   \bigoplus_{s \in I[N]} \scr{B}(\scr{H}^{\otimes 2s}_+).
\end{displaymath}
Here the $\hat Q_{2s}$ are the operations
introduced in Theorem \ref{thm:opt-clo}. Combining $\natpur$ with
$\hat Q$ we get an operation
\begin{equation} \label{eq:29}
  \scr{B}(\scr{H}^{\otimes M}) \ni A \mapsto \optpur(A) := (\natpur \hat Q) (A) \in
  \scr{B}(\scr{H}^{\otimes N})
\end{equation}
which produces, as stated, a fixed number $M$ of
output systems from $N$ input qubits. Physically we can interpret
$\optpur(A)$ in the following way: First we apply the natural
purifier to the input state $\rho(\beta)^{\otimes N}$ and we get $2s$ output
systems in the common state $\rho_s(\beta)$. If $2s \geq M$ we
throw away $M -2s$ qubits and end up with a number of $M$. If $2s
< M$ we have to invoke the $2s \to M$ optimal cloner to reach the
required number of $M$ output systems. Although this cloning
process is wasteful we will see soon that the fidelities
$\scr{F}_\#(\optpur)$ of the output state produced by $\optpur$
are even the best fidelities we can get for any $N \to M$
purifier. Hence we will call $\optpur$ therefore the \emph{optimal
purifier}.

\subsection{The one qubit fidelity} \label{sec:one-qubit-test}

Now we will calculate the one qubit fidelity $\scr{F}_{\rm one}$.
Due to covariance of the depolarizing channel $R$ the
expressions under the infima defining $\scr{F}_{\rm one}(T)$ (and
$\scr{F}_{\rm all}(T)$) in  Equation (\ref{Fall}) and (\ref{Fone})
depend for any fully symmetric purifier not on $\sigma$ and $i$. I.e. we
get with $R_*\sigma = \rho(\beta)$:
\begin{equation} \label{eq:31}
  \scr{F}_{\rm all}(T) = \tr\left[\sigma^{\otimes
M}T_*\bigl(\rho(\beta)^{\otimes
      N}\bigr)\right] \ \mbox{and} \ \scr{F}_{\rm one}(T) =
  \tr\left[\sigma^{(1)}T_*\bigl(\rho(\beta)^{\otimes N}\bigr)\right]
\end{equation} with $\sigma = \kbpp$. In the case of $\scr{F}_{\rm
one}$ the situation is further simplified by the introduction of
the \emph{black cow parameter} (cf. \cite{Klo1}) $\gamma(\theta)$
which is defined for each density matrix $\theta$ on
$\scr{H}^{\otimes M}$ by
\begin{displaymath}
  \gamma(\theta) = \frac{1}{M} \tr(2 L_3 \theta).
\end{displaymath} To derive the relation of $\gamma$ to
$\scr{F}_{\rm one}$ note that full symmetry of $T$ implies
equivalently to (\ref{eq:31}) \begin{displaymath}
  \scr{F}_{\rm one}(T) = \tr\left[ \left(\frac{1}{M} \sum_{j=1}^M
\sigma^{(j)}\right)
    T_*\bigl(\rho(\beta)^{\otimes N}\bigr)\right].
\end{displaymath} Since $\sigma = (\Bbb{1} + \sigma_3)/2$ holds
with the Pauli matrix $\sigma_3$ we get together with the
definition of $L_3$ in Equation (\ref{eq:3}) \begin{equation}
\label{eq:30}
  \scr{F}_{\rm one}(T) = \frac{1}{2}\Bigl[1 +
\gamma\bigl[T_*(\rho(\beta)^{\otimes
    N})\bigr]\Bigr].
\end{equation} In other words it is sufficient to calculate
$\gamma\bigl[T_*(\rho(\beta)^{\otimes N})\bigr]$ (which is simpler
because $\SU(2)$ representation theory is more directly
applicable) instead of $\scr{F}_{\rm one}(T)$.

Another advantage of $\gamma$ is its close relation to the
parameter $\noise = \tanh(\beta)$ defining the operation $R_*$ in
Equation (\ref{eq:1}). In fact we have
\begin{displaymath}
  \gamma(\rho(\beta)^{\otimes N})
    = \frac{1}{N} \tr \bigl(2 L_3 \rho(\beta)^{\otimes N} \bigr)
    = \frac{1}{N} N \tr\bigl(\sigma_3 \rho(\beta)\bigr)
    = \tanh(\beta)
    = \noise.
\end{displaymath}
 In other words the one particle restrictions of the output state
$T\bigl(\rho(\beta)^{\otimes N}\bigr)$ are given by
\begin{displaymath}
  \gamma\bigl[T(\rho(\beta)^{\otimes N})\bigr] \sigma
  + \bigl[1 - \gamma[T(\rho(\beta)^{\otimes N})]\bigr]
  \frac{\Bbb{1}}{2}.
\end{displaymath} This implies that
$\gamma\bigl[T(\rho(\beta)^{\otimes N})\bigr] > \noise$ should
hold if $T$ is really a purifier.

Let us consider now the natural purifier $\natpur$. Since the
number of output qubits is not constant in this case we have to
consider for each $s \in I[N]$ the quantity
 $\scr{F}_{\rm one}(\natpur_s)$ (see Equation (\ref{eq:28})
for the definition of the $\natpur_s$) instead of one fixed
parameter $\scr{F}_{\rm one}(\natpur)$ (in other words: The
fidelity of $\natpur$ is, as the number of output qubits, not a
constant but an observable). According to the discussion above we
get
\begin{align}
  \gamma\bigl(\rho_s(\beta)\bigr)
  &={1\over 2s}\tr\left(2L_3^{(s)} \rho_s(\beta)\right)
  ={1\over 2s}{\tr\bigl(2L_3^{(s)}\exp(2\beta L_3^{(s)})\bigr)\over
    \tr\bigl(\exp(2\beta L_3^{(s)})\bigr)} \notag \\ 
  &={1\over 2s}{d\over d\beta}\ln\tr\bigl(\exp(2\beta L_3^{(s)})\bigr)
  ={1\over 2s}{d\over d\beta}\left(\ln\sinh\bigl((2s+1)\beta\bigr)
    -\ln\sinh\beta\right)\notag  \\
  &={2s+1\over 2s}\coth\bigl((2s+1)\beta\bigr)-{1\over 2s}\coth\beta
  \label{eq:11}
\end{align} and hence
\begin{align}
  \scr{F}_{\rm one}(\natpur_s) &= \frac{1}{2}\Bigl[1 +
  \gamma\bigl(\rho_s(\beta)^{\otimes N}\bigr)\Bigr] \notag \\
  &=  \frac{1}{2}\left[1 + {2s+1\over
    2s}\coth\bigl((2s+1)\beta\bigr)-{1\over 2s}\coth\beta \right] \notag.
\end{align}
 If $s=1/2$ we have
$\gamma\bigl(\rho_s(\beta)\bigr) = \tanh(\beta) = \noise$ hence
the (perturbed) input state $\rho(\beta)$ is reproduced. Taking
the derivative with respect to $s$ shows in addition that
$\gamma\bigl(\rho_s(\beta)\bigr)$ is strictly increasing in $s$.
Hence $\natpur$ really purifies (according to the remark above)
and the best result we get if $s$ is maximal. In the limit $s\to0$
we find $\gamma\bigl(\rho_s(\beta)\bigr) = 0$ which is reasonable
because $\natpur$ does not produce any output at all in this case
($\dim \scr{H}_s = 1$ for $s=0$).

Let us apply these results to the optimal purifier. According to
the definition  of $\optpur$ and $\natpur$ in Equations
(\ref{eq:29}) and (\ref{eq:28}) the decomposition of $\optpur$
given in (\ref{eq:8}) has the form
\begin{equation} \label{eq:32}
  \optpur(A) = \natpur( \hat Q(A))
  = \sum_{s \in I[N]} \hat Q_{2s}(A) \otimes \Bbb{1}
  = \sum_{s \in I[N]} \optpur_s(A) \otimes \Bbb{1},
\end{equation}
 hence $\optpur_s(A) = \hat Q_{2s}(A)$. Together with (\ref{eq:30}) we get
\begin{align}
  \scr{F}_{\rm one}(\optpur)
    &= \frac{1}{2} \left[ 1 + \sum_{s \in I[N]} w_N(s)
        \gamma\bigl[\optpur_{s*} (\rho_s(\beta))\bigr]\right] \label{eq:10}
 \\
    &= \frac{1}{2} \left[ 1 + \sum_{s \in I[N]} w_N(s)
        \gamma\bigl[\hat  Q_{2s*}(\rho_s(\beta))\bigr] \right]  \notag \\
  &=: \sum_{s \in I[N]} w_N(s) f_{\rm one}(M,\beta,s) \notag,
\end{align} where we have introduced the abbreviation
\begin{displaymath}
  f_{\rm one}(M,\beta,s)
   := \frac{1}{2}\left[1 + \gamma\bigl[\hat
         Q_{2s}(\rho_s(\beta))\bigr]\right].
\end{displaymath}
 Together with Theorem \ref{thm:opt-clo} this
implies:
\begin{align*}
  2f_{\rm one}(M,\beta,s) - 1 &=  \gamma\bigl[\hat Q_{2s
*}(\rho_s(\beta))\bigr]
   = \frac{1}{M} \tr\bigl[ 2 \hat Q_{2s}(L_3) \rho_s(\beta) \bigr] \\
  &= \frac{\omega(\hat Q_{2s})}{M} \tr[ 2 L_3^{(s)} \rho_s(\beta) ]
   = \frac{\omega(\hat Q_{2s}) 2s}{M} \gamma[\rho_s(\beta)].
\end{align*}
Inserting the values of $\omega(\hat Q_{2s})$ and
$\gamma[\rho_s(\beta)]$ from Equations (\ref{eq:25}) and
(\ref{eq:11}) we get
\begin{multline}  \label{eq:36}
  2 f_{\rm one}(M,\beta,s) - 1 = \\
  =\begin{cases}
    \displaystyle {2s+1\over 2s}\coth\bigl((2s+1)\beta\bigr)-{1\over
      2s}\coth \beta & \mbox{for $2s > M$} \\[10pt]
    \displaystyle \frac{1}{2s+2} \frac{M+2}{M}
      \Bigl((2s+1)\coth\bigl((2s+1)\beta\bigr) - \coth\beta\Bigr) &
      \mbox{for $2s \leq M$.}
  \end{cases}
\end{multline} Hence we have proved the following proposition.

\begin{prop}
  The one--qubit fidelity $\scr{F}_{\rm one}(\optpur)$ of the
  optimal purifier is given by
  \begin{equation} \label{eq:18}
    \scr{F}_{\rm one}(\optpur) = \sum_{s \in I[N]} w_N(s) f_{\rm
      one}(M,\beta,s)
  \end{equation}
  with $f_{\rm one}(M,\beta,s)$ from Equation (\ref{eq:36}).
\end{prop}

Note in particular that in the case $M = 1$ the one--qubit
fidelity coincides with the expectation value of the fidelity of
$\natpur$ in the state $\natpur_*(\rho(\beta)^{\otimes N})$ -- the
\emph{mean fidelity}. Hence we can reinterpret the natural
purifier as a device which produces exactly one output system (cf.
\cite{IAC}).

\subsection{The all qubit fidelity} \label{sec:all-qubit-test}

As in the one--qubit case the all--qubit fidelity of $\natpur$ is
an observable rather than a fixed parameter. Hence we have to
calculate $\scr{F}_{\rm all}(\natpur_s)$ for each fixed $s$.
Applying again Equation (\ref{eq:31}) we get
\begin{align}
  \scr{F}_{\rm all}(\natpur_s)
    &= \tr \bigl(\sigma^{\otimes 2s} \rho_s(\beta)\bigr)
     = \frac{\sinh(\beta)}{\sinh\bigl((2s+1)\beta\bigr)} e^{2\beta s} \notag \\
    &= \frac{e^{(2s+1)\beta} - e^{(2s-1)\beta}}{e^{(2s+1)\beta} - e^{(2s+1)\beta}}
    = \frac{1 - e^{-2\beta}}{1 - e^{-(4s+2)\beta}}. \notag 
\end{align}

Using the decomposition of $\optpur$ given in Equation
(\ref{eq:32}) we get for the optimal purifier something similar as
in the last subsection:
\begin{align} \label{eq:16}
  \scr{F}_{\rm all}(\optpur)
  &=  \sum_{s \in I[N]} w_N(s) \tr\left[\sigma^{\otimes M} \optpur_{s*}
    \bigl(\rho_s(\beta)\bigr)\right] \\
  &= \sum_{s \in I[N]} w_N(s) \tr\left[\sigma^{\otimes M}\hat
    Q_{2s*}\bigl(\rho_s(\beta)\bigr)\right]. \notag
\end{align} However the calculation of
\begin{displaymath}
   f_{\rm all}(M,\beta,s) := \tr\left[\sigma^{\otimes M}\hat
     Q_{2s*}\bigl(\rho_s(\beta)\bigr)\right]
\end{displaymath} is now more difficult, since the knowledge of
$\hat Q_{2s}(L_3) = \omega(\hat Q_{2s}) L_3^s$ is not sufficient in this
case. Hence we have to use the explicit form of $\hat Q_{2s}$ in
Equation (\ref{eq:24}) and (\ref{eq:26}). For $2s < M$ this leads to
\begin{eqnarray}
  f_{\rm all}(M,\beta,s)
  &=&\frac{2s+1}{M+1}\ \langle\psi^{\otimes M},
s_M(\rho_s\otimes\idty^{\otimes (M-2s)})
       s_M\psi^{\otimes M}\rangle
  \nonumber\\
  &=&\frac{2s+1}{M+1}\ \langle\psi^{\otimes M}, (\rho_s\otimes\idty^{\otimes(M-2s)})
       \psi^{\otimes M}\rangle
   = \frac{2s+1}{M+1}\ \langle\psi^{\otimes 2s},
        \rho_s\psi^{\otimes 2s}\rangle\nonumber\\
  &=&\frac{2s+1}{M+1}\ \frac{1-e^{-2\beta}}{1-e^{-(4s+2)\beta}}.
\nonumber 
\end{eqnarray}

For $M\leq2s$ we have to calculate
\begin{align}
  f_{\rm all}(s,M,\beta) &= \tr\left[\sigma^{\otimes M}\hat
    Q_{2s*}\bigl(\rho_s(\beta)\bigr)\right]
  = \tr\left[\hat Q_{2s}(\sigma^{\otimes M})
        \rho_s(\beta)\right] \notag \\
  &= \tr\left[\rho_s(\beta)\Bigl(S_M[(\vert\psi^{\otimes M}\rangle\langle\psi^{\otimes
      M}\vert)\otimes\idty^{\otimes(2s-M)}]S_M\Bigr)\right]
  \label{fidall:b}
\end{align}
 We will compute the operator $\hat Q_{2s}(\sigma^{\otimes M})$ in
occupation number representation. By definition, the basis vector
``$\vert n\rangle$'' of the occupation number basis is the
normalized version of  $S_M\Psi$, where $\Psi$ is a tensor product
of $n$ factors $\psi$ and $(M-n)$ factors $\phi$, where $\phi={ 0
\choose 1}$ denotes obviously the second basis vector. The
normalization factor is easily computed to be
\begin{equation}\label{occbasis}
  S_M (\psi^{\otimes n}\otimes\phi^{\otimes(M-n)})
  ={M\choose n}^{-1/2} \vert n\rangle.
\end{equation} We can now expand the ``$\idty$'' in Equation
(\ref{fidall:b}) in product basis, and apply (\ref{occbasis}), to
find
\begin{displaymath}
  S_M[(\vert\psi^{\otimes M}\rangle\langle\phi^{\otimes
M}\vert)\otimes\idty^{\otimes(2s-M)}]S_M
  =\sum_K{2s-M\choose K-M}{2s\choose K}^{-1}\
  \vert K\rangle\,\langle K\vert.
\end{displaymath} Now $L_3$ is diagonal in this basis, with
eigenvalues $m_K=(K-s)$, $K=0,\ldots,(2s)$. With $\rho_s(\beta)$
from (\ref{eq:14}) we get
\begin{displaymath}
  f_{\rm all}(M,\beta,s)
  =\frac{1-e^{-2\beta}}{1-e^{-(4s+2)\beta}}
  \sum_K{2s-M\choose K-M}{2s\choose K}^{-1}\ e^{2\beta(K-s)}
  \quad\text{for $M\leq 2s$}.
\end{displaymath} Together with \begin{displaymath}
  {2s-M\choose K-M}{2s\choose K}^{-1} = \frac{(2s-M)!}{ (K-M)!
    (2s-K)!} \frac{K!(2s-K)!}{(2s)!} = {2s \choose M}^{-1} {K
    \choose M}
\end{displaymath} we get
\begin{displaymath} 
  f_{\rm all}(M,\beta,s) = \frac{1-e^{-2\beta}}{1-e^{-(4s+2)\beta}} {2s
    \choose M}^{-1}\sum_K{K\choose M} e^{2\beta(K-s)}.
\end{displaymath} Summarizing these calculations we get the
following proposition:

\begin{prop}
  The all--qubit fidelity $\scr{F}_{\rm all}(\optpur)$ of the
  optimal purifier is given by
  \begin{equation} \label{eq:21}
    \scr{F}_{\rm all}(\optpur) =  \sum_{s \in I[N]} w_N(s) f_{\rm
one}(M,\beta,s)
  \end{equation}
  where $f_{\rm all}(M,\beta,s)$ is given by
  \begin{multline} \label{eq:4}
    f_{\rm all}(M,\beta,s) =
    \begin{cases}
      \displaystyle \frac{2s+1}{M+1}\
\frac{1-e^{-2\beta}}{1-e^{-(4s+2)\beta}}
      & \mbox{$M \leq 2s$} \\[10pt]
      \displaystyle \frac{1-e^{-2\beta}}{1-e^{-(4s+2)\beta}} {2s \choose
        M}^{-1}\sum_K{K\choose M} e^{2\beta(K-s)} & \mbox{$M > 2s$.}
    \end{cases}
  \end{multline}
\end{prop}

\section{Solution of the optimization problems}
\label{sec:Solut-optim-probl}

Now we are going to prove the following theorem:

\begin{thm}
  The purifier $\optpur$ maximizes the fidelities $\scr{F}_{\rm
    one}(T)$ and $\scr{F}_{\rm all}(T)$. Hence the optimal fidelities
  $\fidopt_{\rm one}(N,M)$ and $\fidopt_{\rm all}(N,M)$ defined in
  Section \ref{sec:merit} are given by Equation (\ref{eq:18}) and
  (\ref{eq:21}).
\end{thm}

\begin{proof} Note first that the funtionals $\scr{F}_{\rm one}$
and $\scr{F}_{\rm all}$ are, as infima over continuous functions,
upper semicontinuous. Together with the compactness of the set of
admissible $T$ this implies that the suprema $\fidopt_\#(N,M)$
from Equation (\ref{Fopt}) are attained. In other words: optimal
purifier $T$ with $\scr{F}_\#(T) = \fidopt_\#(N,M)$ exist, and we
can assume without loss of generality that they are fully
symmetric (according to the discussion in
Section~\ref{sec:decomp-reduc}). Hence we can apply Equation
(\ref{eq:31}) and the decomposition (\ref{eq:8}) to get in analogy
to (\ref{eq:10}) and (\ref{eq:16})
\begin{displaymath}
 \scr{F}_{\rm one}(T) = \frac{1}{2} \left[ 1 + \sum_{s \in I[N]}
    w_N(s) \gamma\bigl[T_{s*} (\rho_s(\beta))\bigr]\right]
\end{displaymath} and
\begin{equation} \label{eq:38}
  \scr{F}_{\rm all}(T)
    =\sum_{s \in I[N]} w_N(s) \tr\left[\sigma^{\otimes M} T_{s*}
         \bigl(\rho_s(\beta)\bigr)\right].
\end{equation}

The last two Equations show that we have to optimize each
component $T_s$ of the purifier $T$ independently. In the one
qubit case this is very easy, because we can use Theorem
\ref{thm:opt-clo} to get $T_s(L_3) = \omega(T_s)L_3^{(s)}$ and
$\gamma\bigl[T_{s*}(\rho_s(\beta))\bigr] = \omega(T_s)
\tr\bigl(L_3^{(s)}\rho_s(\beta)\bigr)$. Hence maximizing
$\gamma\bigl[T_{s*}(\rho_s(\beta)\bigr)]$ is equivalent to
maximizing $\omega(T_s)$. But we have according to Theorem
\ref{thm:opt-clo}
  \begin{displaymath}
    \max_T \omega(T_s) = \omega(\hat Q_{2s}) =
    \begin{cases}
      \displaystyle \frac{M}{2s} & \mbox{for $2s \geq M$} \\[10pt]
      \displaystyle \frac{M+2}{2(s+1)} & \mbox{for $2s < M$},
    \end{cases}
  \end{displaymath}
  which shows that $\fidopt_{\rm one}(N,M) = \scr{F}_{\rm one}(\optpur)$
  holds as stated.

For the many qubit--test version the proof is slightly more
difficult. However as in the $\scr{F}_{one}$-case we can solve the
optimization problem for each summand in Equation (\ref{eq:38})
separately. First of all this means that we can assume without
loss of generality that $T_{s*}$ takes its values in
$\scr{B}(\scr{H}_+^{\otimes M})$ because the functional
\begin{equation} \label{eq:22}
  f_s(T_s) := \tr\left(\sigma^{\otimes
M}T_{s*}\bigl(\rho_s(\beta)\bigr)\right)
\end{equation} which we have to maximize, depends only on this
part of the operation. Full symmetry implies in addition that
$T_{s*}(\rho_s(\beta))$ is diagonal in occupation number basis
(see Equation (\ref{occbasis})), because $T_{s*}(\rho_s(\beta))$
commutes with each $\pi_{s'}(U)$ ($s' = M/2$, $U \in \U(2)$) if
$\pi_s(U)$ commutes with $\rho_s(\beta)$.

If $M > 2s$ this means we have $T_{s*}(\rho_s(\beta)) = \kappa_*
\sigma^{\otimes M} + r_*$ where $r_*$ is a positive operator with
$\sigma^{\otimes M} r_* = r_* \sigma^{\otimes M} = 0$. Inserting
this into (\ref{eq:22}) we see that $f_s(T_s) = \kappa_*$. Hence we
have to maximize $\kappa_*$. The first step is an upper bound
which we get from the fact that $\tr\bigl(\sigma^{\otimes M}
\rho_s(\beta)\bigr) \Bbb{1} - \rho_s(\beta)$ is a positive
operator. Since $T_{s*}(\Bbb{1}) = (2s+1)/(M+1) \Bbb{1}$ (another
consequence of full symmetry) we have \begin{displaymath}
  0 \leq T\Bigl(\tr\bigl(\sigma^{\otimes2s} \rho_s(\beta)\bigr) \Bbb{1} -
\rho_s(\beta)\Bigr) =
  \frac{2s+1}{M+1} \tr\bigl(\sigma^{\otimes M} \rho_s(\beta)\bigr) \Bbb{1}
- \kappa \sigma^{\otimes M} -
  r_*.
\end{displaymath} Multiplying this Equation with $\sigma^{\otimes
M}$ and taking the trace we get \begin{equation}\label{eq:23}
  \kappa_* \leq \frac{2s+1}{M+1} \tr\bigl(\sigma^{\otimes M}
\rho_s(\beta)\bigr).
\end{equation} However calculating $f_s(\optpur_s)$ we see that
this upper bound is achieved, in other words $\optpur_s$ maximizes
$f_s$.

If $M \leq 2s$ holds we have to use slightly different arguments
because the estimate (\ref{eq:23}) is to weak in this case.
However we can consider in Equation (\ref{eq:22}) the dual $T_s$
instead of $T_{s*}$ and use then similar arguments. In fact for
each covariant $T_s$ the quantity $T_s(\sigma^{\otimes M})$ is,
due to the same reasons as $T_{s*}(\rho_s(\beta))$ diagonal in the
occupation number basis and we get $T_s(\sigma^{\otimes M}) =
\kappa \sigma^{\otimes2s} + r$ where $r$ is again a positive
operator with $r = \sum_{n=0}^{2s-1} r_n |n\rangle$ ($|n\rangle$
denotes again the occupation number basis) and $\kappa$ is a
positive constant. Since $T_s$ is unital we get from $\Bbb{1} -
\sigma^{\otimes M} \geq 0$ the estimate $0 \leq \kappa \leq 1$ in
the same way as Equation (\ref{eq:23}). Calculating $\optpur_s(\sigma^{\otimes
  M})$ shows again that the upper bound $\kappa = 1$
is indeed
achieved, however it is now not clear whether maximizing $\kappa$
is equivalent to maximizing $f_s(T_s)$.

Hence let us show first that $\kappa = 1$ is \emph{necessary} for
$f_s(T_s)$ to be maximal. This follows basically from the fact
that $T_s$ is, up to a multiplicative constant, trace preserving.
In fact we have \begin{displaymath}
  \tr\bigl(T_s(\sigma^{\otimes M})\bigr) = \tr\bigl(T_s(\sigma^{\otimes
M})\Bbb{1}\bigr) =
  \tr\bigl( \sigma^{\otimes M} T_{s*}(\Bbb{1}) \bigr) = \frac{2s+1}{M+1}.
\end{displaymath} This means especially that $\kappa + \tr(r) =
(2s+1)/(M+1)$ holds, i.e. decreasing $\kappa$ by $0 < \epsilon <
1$ is equivalent to increasing $\tr(r)$ by the same $\epsilon$.
Taking into account that $\rho_s(\beta) = \sum_{n=0}^{2s} h_n
|n\rangle$ holds with $h_n =
\exp\bigl(2\beta(n-s)\bigr)$, we see that reducing $\kappa$ by
$\epsilon$ reduces $f_s(T_s)$ at least by \begin{displaymath}
  \epsilon \Bigl( \tr\bigl(\sigma^{\otimes2s} \rho_s(\beta)\bigr) -
\tr\bigl(|2s-1\rangle
  \rho_s(\beta)\bigr)\Bigr) = \epsilon \bigl(e^{2\beta s} -
e^{(2s-1)\beta}\bigr) > 0.
\end{displaymath} Therefore $\kappa = 1$ is necessary.

The last question we have to answer, is how the rest term $r$ has
to be chosen, for $f_s(T_s)$ to be maximal. To this end let us
consider the slightly modified fidelity  $\tilde f_s(T_s) =
\tr\bigl(T_s(\sigma^{\otimes M}) \sigma^{\otimes2s}\bigr)$ (which
is in fact related to optimal cloning; see \cite{Klo1} and Section
\ref{sec:decomp-cpmaps}). It is in contrast to $f_s(T_s)$
maximized \emph{iff} $\kappa = 1$. However the operation which
maximizes $\tilde f(T_s)$ is obviously the optimal $M \to 2s$
cloner (up to normalization) which is according to \cite{Klo2}
unique. This implies that $\kappa=1$ fixes $T_s$  already.
Together with the facts that $\kappa=1$ is necessary for
$f_s(T_s)$ to be maximal and $\kappa=1$ is realized for
$\optpur_s$ we conclude that $\max f_s(T_s) = f_s(\optpur_s)$
holds, which proves the assertion.
\end{proof}

\section{Asymptotic behaviour}
\label{sec:asympt}

Now we want to analyze the rate with which nearly perfect purified
qubits can be produced in the limit $N\to\infty$. To this end we
have to compute the asymptotic behaviour of various expectations
involving $s$. It turns out that it is much better not to do work
with the explicit expressions of these expectations, as sums over
expressions with many binomial coefficients, but to go back to the
definition, and use general properties of expectations of
$\rho^{\otimes N}$. This has the added advantage of being easily
generalized to Hilbert space dimensions $d>2$, so we expect the
method to be useful in its own right. We collect the basic
statements in the following subsection, applying them to the
concrete expressions in subsequent ones.

\subsection{Convergence of weights to a point measure}
\label{sec:asy-point}

In the classical case the general theory alluded to above is
nothing but the theory of asymptotic distributions for independent
identically distributed random variables (Laws of large numbers of
various sorts). In the quantum case this theory has been developed
in the context of the statistical mechanics of general mean-field
systems \cite{MF}. Of this theory we need only the simplest
aspects (convergence to a point measure), and not the more
advanced ``Large Deviation'' parts, in which it is shown how the
probability of deviations from the limit decrease exponentially
fast.

Consider operators of the form $A_N=(1/N)\sum_{i=1}^N a^{(i)}$,
where $a^{(i)}$ denotes the copies of a fixed operator on
$\scr{H}$, acting in the $i^{\rm th}$ tensor factor of
$\scr{H}^{\otimes N}$. It is clear that the expectations
$\tr(\rho^{\otimes N}A_N)=\tr(\rho a)$ are independent of $N$. Now
consider products of a finite number of such operators and expand
the expectation into the average over all terms of the form
 $\tr(\rho^{\otimes N}a^{(i)}b^{(j)}c^{(k)}\cdots)$.
It is easy to see that for large $N$ the majority of these terms
will be such that all indices $i,j,k,\ldots$ are different, and
for such terms the above expression is equal to
 $\tr(\rho a)\tr(\rho b)\tr(\rho c)\cdots$.
So this will be the limit of the expectation of the product
$A_NB_NC_N\cdots$ as $N\to\infty$ (for precise combinatorial
estimates, see \cite{MF}). Of course, this allows us to compute
the asymptotic expectations for arbitrary polynomials, and by
taking suitable limits of arbitrary continuous functions of
Hermitian operators. There is an abstract non-commutative
functional calculus describing exactly these possibilities (see
appendix of \cite{MF}). However, for our purposes it is sufficient
to say that all combinations of algebraic operations and
continuous functions of a Hermitian variable (evaluated in the
usual spectral functional calculus) are in this class.

For the case at hand, note that the angular momentum operators
$L_k$ as in Equation (\ref{eq:3}) are of the form $NA_N$ therefore,
for any sequence of functions $f_N$ of three non-commuting
arguments (this means that in writing out $f_N$ we have to keep
track of operator ordering), which converges to a limit function,
$f_\infty$, we get
\begin{equation} \label{MFlim}
  \lim_{N\to\infty}\tr\left(\rho^{\otimes N}
    f_N\bigl({\textstyle{L_1\over N},{L_2\over N},{L_3\over N}}\bigr)\right)
  =f_\infty\left(\tr(\rho {\sigma_1\over2}),\tr(\rho
    {\sigma_2\over2}),\tr(\rho {\sigma_3\over2})\right).
\end{equation}
 Note that the function $f_\infty$ is just evaluated on numbers
(operators on a one-dimensional space) so all operator ordering
problems disappear in the limit. This is the huge simplification
which makes mean-field theory so accessible. The limit formula
will be applied to functions of ``$2s$'', the number of outputs
from the natural purifier, which can itself be written as a
function of this sort. It is, of course, constant on each summand
of the decomposition (\ref{eq:pidecomp}), so it is a function of
the Casimir operator $\vec L^2=s(s+1)$:
\begin{align}
    {2s\over N}=g_N\bigl({\textstyle{L_1\over N},{L_2\over
        N},{L_3\over N}}\bigr)
        &=\sqrt{4(\vec L/N)^2+N^{-2}}-1/N \notag     \\
    g_\infty(x_1,x_2,x_3) &=\lim_{N\to\infty}\sqrt{4(\vec x)^2+N^{-2}}-1/N=
    2 |\vec{x}| \label{fN}   \notag\\
    g_\infty\left(\tr(\rho {\sigma_1\over2}),\tr(\rho
    {\sigma_2\over2}),\tr(\rho {\sigma_3\over2})\right)
       &= g_\infty(0,0,\noise/2)=\noise=\tanh\beta,
\end{align}
 when $\rho=\rho(\beta)$ is given by eq.(\ref{eq:14}).
Functions of $g$ then also lie in the relevant functional
calculus, so we get the following statement, taylored to our need
in the following subsections. In it we have already encorporated
further, straightforward approximation arguments, using uniformly
convergent sequences of continuous functions to establish upper
and lower bounds separately.

\begin{lem} \label{thm:ptmeas}
  Let  $f_N: (0,1) \to \Bbb{R}$, $N \in \Bbb{N}$ be a uniformly bounded
sequence of
continuous functions, converging uniformly on a neighborhood of
$\noise = \tr(\rho(\beta) \sigma_3)$ to a continuous function
$f_\infty$, and let $w_N(s)$ denote the weights in Equation
(\ref{eq:2}). Then
 \begin{equation}
  \lim_{N\to\infty} \sum_{s \in I[N]} w_N(s) f_N(2s/N)
      = f_\infty(\noise).
 \end{equation} \end{lem}

\begin{figure}[htbp]
  \begin{center}
    \includegraphics{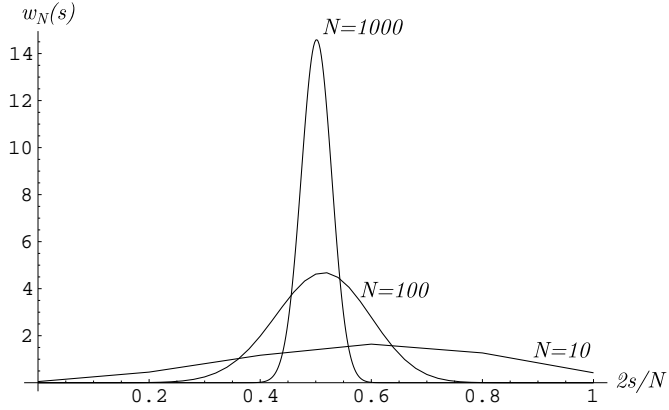}
    \caption{Convergence of $w_N(s)$ to a point measure
       ($\noise=.5$, $N=10,100,1000$).\newline
       Discrete points joined, and rescaled for total area $1$}
    \label{fig:ptmeas}
  \end{center}
\end{figure}

In the language of measure theory this is saying that the
probability measures $\sum_{s} w_N(s) \delta(x-2s/N)dx$ on the
interval $[0,1]$ converge to the point measure
$\delta(x-\lambda)dx$. Graphically, this is shown in
Figure~\ref{fig:ptmeas}

\subsection{The one particle test}
\label{sec:asympt-one}

Let us analyze first the behaviour of the optimal one--qubit
fidelity $\fidopt_{\rm one}(N,M)$ in the limit $M\to\infty$.
Obviously only the $M>2s$ case of $f_{\rm one}(M,\beta,s)$ is
relevant in this situation and we get, together with Equation
(\ref{eq:18}), the expression
\begin{multline*} 
  \fidopt_{\rm one}(N,\infty) = \\ \sum_{s \in I[N]} w_N(s) \frac{1}{2}
\left[ 1
    + \frac{1}{2s+2} \Bigl((2s+1)\coth\bigl((2s+1)\beta\bigr)
    - \coth\beta\Bigr) \right],
\end{multline*}
which obviously takes its values between $0$ and $1$. To take the
limit $N\to\infty$ we can write
\begin{displaymath}
  \lim_{N\to\infty} \fidopt_{\rm one}(N,\infty) = \lim_{N\to\infty}
\sum_{s\in I[N]} w_N(s)
  f_{N,\infty}(\frac{2s}{N})
\end{displaymath}
with
\begin{displaymath}
  f_{N,\infty}(x) = \frac{1}{2} \left[ 1 + \frac{1}{Nx+2}
    \Bigl((Nx+1)\coth\bigl((Nx+1)\beta\bigr) - \coth\beta\Bigr)\right].
\end{displaymath}
The functions $f_{N,\infty}$ are continuous, bounded and converge
on each interval $(\epsilon,1)$ with $0< \epsilon < 1$ uniformly
to $f_{\infty,\infty} \equiv 1$. Hence the assumptions of
Lemma~\ref{thm:ptmeas} are fulfilled and we get
\begin{displaymath}
  \lim_{N\to\infty} \fidopt_{\rm one}(N,\infty) = f_{\infty,\infty}(\noise)
= 1
\end{displaymath}
as already stated in Section \ref{sec:merit}. This means that we
can produce arbitrarily good purified qubits at infinite rate if
we have enough input systems.

To analyze how fast the quantity $\fidopt_{\rm one}(N,\infty)$
approaches 1 as $N\to\infty$ let us consider the limit
\begin{equation} \label{eq:5}
   \lim_{N\to\infty} N (1 - \fidopt_{\rm one}(N,\infty))
   = \sum_{s \in I[N]} w_N(s)\tilde f_{N,\infty}(\frac{2s}{N})
   \equiv \frac{c_\infty}{2}
\end{equation}
with $\tilde f_{N,\infty} = N(1 - f_{N,\infty})$. The existence of
this limit is equivalent to the asymptotic formula
\begin{displaymath}
  \fidopt_{\rm one}(N,\infty) = 1 - \frac{c_\infty}{2N} +
\order{\frac{1}{N}}
  ,
\end{displaymath}
 where, as usual, $\order{\frac{1}{N}}$ stands for terms going to
zero faster than $\frac{1}{N}$. Lemma~\ref{thm:ptmeas} leads to
$c_\infty/2
   = \tilde f_{\infty,\infty}(\lambda)$ with
$\tilde f_{\infty,\infty} = \lim_{N \to \infty} \tilde
f_{N,\infty}$ uniformly on $(\epsilon,1)$. To calculate $\tilde
f_{\infty,\infty}$ note that
\begin{displaymath}
  \tilde f_{N,\infty} (x) = \frac{N}{Nx +2} + \frac{N \coth \beta}{Nx +2} +
\
  \mbox{Rest}
\end{displaymath}
holds, where ``Rest'' is a term which vanishes exponentially fast
as $N\to\infty$. Hence with $\coth \beta = 1 / \noise$ we get
\begin{displaymath}
  c_\infty = 2 \tilde f_{\infty,\infty}(\noise) = \frac{1 +
\noise}{\noise^2}
\end{displaymath}

The asymptotic behaviour of $\fidopt_{\rm one}(N,1)$ can be analyzed
in the same way. The only difference is that we have to consider now
the $1 = M \leq 2s$ branch of Equation (\ref{eq:36}). In analogy to
Equation (\ref{eq:5}) we have to look at
\begin{displaymath}
   \lim_{N\to\infty} N (1 - \fidopt_{\rm one}(N,1)) = \sum_{s \in I[N]}
w_N(s)
   \tilde f_{N,1}(\frac{2s}{N}) = \frac{c_1}{2}
\end{displaymath}
with $\tilde f_{N,1} = N(1 - f_{N,1})$ and
\begin{displaymath}
 f_{N,1}(x) = \frac{1}{2} \left[ 1- \frac{1}{Nx} \left[
     (Nx+1)\coth\bigl((Nx+1)\beta\bigr)-\coth \beta \right] \right].
\end{displaymath}
For $\tilde f_{\infty,1}$ we get
\begin{equation} \label{eq:6}
  \tilde f_{\infty,1}(x) = \frac{1}{2} (\frac{-1}{x} + \frac{1}{x \noise}).
\end{equation}
Using again Lemma~\ref{thm:ptmeas} leads to
\begin{displaymath}
  c_1 = 2 \tilde f_{\infty,1}(\noise) = \frac{1-\noise}{\noise^2}.
\end{displaymath}

Finally let us consider $\fidopt_{\rm one}(N,0)$. Here the
situation is easier than in the other cases because $\fidopt_{\rm
one}(N,0)$ equals the fidelity of the best possible output of the
natural purifier, i.e.
\begin{displaymath}
  \fidopt_{\rm one}(N,0) = \frac{1}{2} \left[ 1- \frac{1}{N}
    \left[(N+1)\coth\bigl((N+1)\beta\bigr)-\coth \beta \right] \right] =
  f_{N,1}(1).
\end{displaymath}
Hence we only need the asymptotic behaviour of $f_{N,1}(x)$ at
$x=1$. Using Equation (\ref{eq:6}) we get
\begin{displaymath}
  \fidopt_{\rm one}(N,0) = 1 - \frac{1-\noise}{\noise} \frac{1}{2N} +
  \cdots .
\end{displaymath}
This concludes the proof of Equations (\ref{eq:cs}) to
(\ref{eq:ci}).

\subsection{The many particle test}
\label{sec:asympt-all}

Consider now the many--qubit fidelity ${\scr F}_{\rm all}$.
Although, like $\scr{F}_{\rm one}$, it lies between zero and one,
and would attain the value $1$ precisely for a (non-existent)
ideal purifier, both quantities behave quite differently, when we
use them to compare states in systems of varying size. We are
looking here at the two kinds of fidelities for an $M$-particle
output state $\rho_M$ with respect to a one-particle pure state
given by the vector $\psi$, namely
\begin{eqnarray}
  F_{\rm all}
    &=&\langle\psi^{\otimes M},\rho_M\psi^{\otimes M}\rangle
    =\tr\rho_M\Bigl(\vert\psi^{\otimes M}\rangle
                  \langle\psi^{\otimes M}\vert\Bigr)
\quad\text{, and}\nonumber\\
  F_{i}&=&\langle\psi,\rho_M^{(i)}\psi\rangle
       =\tr\rho_M\Bigl(\idty\otimes\cdots
          (\vert\psi\rangle\langle\psi\vert)_i\otimes\idty\Bigr)
\quad,\nonumber
\end{eqnarray}
where $\rho_M^{(i)}$ denotes the restriction of $\rho_M$ to the
$i^{\rm th}$ tensor factor. Let $p_{\rm all}$ and $p_i$ denote the
projections whose $\rho_M$-expectations appear on the right hand
side of these Equations. These projections commute, and $p_{\rm
all}$ is the intersection (in the commuting case: the product) of
the $p_i$ in the lattice of projections. This corresponds to the
union of the respective complements, i.e.,
\begin{displaymath}
   \idty-p_i\leq \idty -p_{\rm all} \leq \sum_i(\idty-p_i)
\quad.
\end{displaymath}
Taking expectations with respect to $\rho_M$, we find that
$\sup_i(1-F_i)\leq(1-F_{\rm all})\leq\sum_i(1-F_i)\leq
M\sup_i(1-F_i)$. For the two figures of merit introduced in
Section~\ref{sec:intro} this implies
\begin{equation}\label{Fone<>Fall}
  (1-{\scr F}_{\rm one}(T))
     \leq (1- {\scr F}_{\rm all}(T))
     \leq M(1-{\scr F}_{\rm one}(T))
\quad,
\end{equation}
for every purifying device $T$. Hence, for fixed $N$ the two
figures of merit are equivalent to within a factor . But the upper
bound becomes meaningless in the limit $M\to\infty$, so it is not
clear at all whether we can bring the fidelity ${\scr F}_{\rm
all}(T)$ close to one for an increasing number of outputs.

As a consequence of this analysis it is necessary to perform the limit
$N,M\to\infty$ more carefully as in the one qubit case. We will consider
therefore the limits $N\to\infty$ and $M\to\infty$ simultaneously, while the
quotient $M/N$ approaches a constant $\rate$, i.e. we will calculate the
function $\Phi(\rate)$ defined in Equation (\ref{eq:20}). The first step in
this context is the following lemma, which allows us to handle the
${2s \choose M}^{-1}\sum_K{K\choose M} e^{2\beta(K-s)}$ term in Equation
(\ref{eq:4}).

\begin{lem} \label{lem:binom}
  For integers $M\leq K$ and $z\in\Bbb{C}$, define
  \begin{displaymath}
    \Phi(K,M,z)={K\choose M}^{-1}\ \sum_{R=M}^K {R\choose M} z^{K-R}.
  \end{displaymath}
  Then, for $|z|<1$, and $c\geq1$:
  \begin{displaymath}
    \lim_{M,K\to\infty\atop M/K \to c} \Phi(K,M,z)={1\over 1-(1-c)z}.
  \end{displaymath}
\end{lem}

\begin{proof}
  We substitute $R\mapsto(K-R)$ in the sum, and get
  \begin{displaymath}
    \Phi(K,M,z)= \sum_{R=0}^\infty c(K,M,R) z^{R},
  \end{displaymath}
  where coefficients with  $M+R>K$ are defined to be zero. We can
  write the non-zero coefficients as
  \begin{align}
    c(K,M,R)&={K\choose M}^{-1}\ {K-R\choose M}
    ={(K-M)! (K-R)!\over K! (K-R-M)!}\notag \\
    &={(K-M)\over K}{(K-M-1)\over(K-1)}\cdots
    {(K-M-R+1)\over(K-R+1)}\notag \\
    &=\prod_{S=0}^{R-1}\Bigl(1-{M\over K-S} \Bigr) \notag.
  \end{align}
  Since $0\leq c(K,M,R)\leq1$,  for all $K,M,R$, the series for
  different values of $M,K$ are all dominated by the geometric
  series, and we can go to the limit termwise, for every $R$
  separately. In this limit we have $M/(K-S)\to c$ for every $S$,
  and hence $c(K,M,R)\to (1-c)^R$. The limit series is again
  geometric, with quotient $(1-c)z$ and we get the result.
\end{proof}

To calculate now $\Phi(\rate)$ recall that  the weights $w_N(s)$
approach a point measure in $2s/N =: x$ concentrated at $\noise =
\tr(\rho(\beta)\sigma_3)$. This means that in Equation
(\ref{eq:21}) only the term with $2s = \noise N$ survives the
limit. Hence if $\rate \geq \noise$ we get $M \geq \noise N = 2s$.
Using Equation (\ref{eq:4}) and Lemma~\ref{thm:ptmeas} we get in
this case
\begin{displaymath}
  \Phi(\rate) = \frac{\noise}{\rate} (1-e^{-2\beta}).
\end{displaymath}
We see that $\Phi(\rate) \to 0$ for $\rate \to \infty$ and $\Phi(\rate) \to
1-\exp(-2\beta)$ for $\rate \to \noise$.

If $0 < \rate  < \noise$ we get $M < \noise N = 2s$, which means
we have to choose Equation (\ref{eq:4}) for
 $f_{\rm all}(M,\beta,s)$. With Lemma \ref{lem:binom} and
Lemma~\ref{thm:ptmeas} we get
\begin{displaymath}
  \Phi(\rate) = \frac{1-e^{-2\beta}}{1 - (1-\rate/\noise)e^{-2\beta}}
\end{displaymath}
which approaches $1$ if $\rate \to 0$ and $1-\exp(-2\beta)$ if
$\rate \to \noise$. Writing this in terms of $\noise=\tanh\beta$,
we obtain Equation (\ref{eq:20}).

\subsection{Estimating the many particle fidelity in terms of one particle }
\label{sec:asympt-est}

\begin{figure}[htbp]
  \begin{center}
    \includegraphics[scale=1]{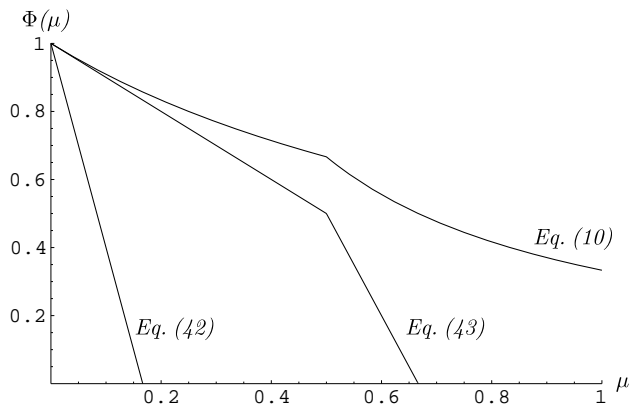}
    \caption{The lower bounds (\ref{eq:phicrude}) and (\ref{eq:phibelow})
    together with the exact result (\ref{eq:20}) \newline
    for the all-particle test fidelity as a function of the rate
($\noise=.5$)}
    \label{fig:phibound}
  \end{center}
\end{figure}

In Section~\ref{sec:merit} we motivated the observation that the
the best all-particle fidelity is a function of the rate (and not
identically equal to $1$) by estimating the all-particle fidelity
in terms of the one-particle fidelity. Since the latter quantity
tends to be more easily computable it is of some interest for
further investigations, how good that estimate actually is. The
estimate mentioned in the text before Equation (\ref{eq:20})
amounts to
\begin{equation}\label{eq:phicrude}
  \Phi(\rate)
    \geq 1- \frac{\rate}{2}c_\infty
    =    1- \frac{\rate(\noise+1)}{2\noise}.
\end{equation}
However, the same basic estimate via  Equation (\ref{Fone<>Fall})
gives even more information:
\begin{align}
 \Phi(\rate)&\geq 1- \lim_{N\to\infty\atop M/N\to\rate}\
                     M(1-\fidopt_{\rm one}(N,M))\notag\\
            &\geq 1- \rate\lim_{N\to\infty}\
                    \sum_{s \in I[N]} w_N(s)\
                     N(1-f_{\rm one}(\rate N,\beta,s))\notag\\
            &= \begin{cases}
            \displaystyle 1-\frac{\rate(1-\noise)}{2\noise^2} &
                            \mbox{if $\rate \leq \noise$}\\
            \displaystyle2-\frac{\rate(1+\noise)}{2\noise^2} &
                           \mbox{if $\rate \geq \noise$,}\\
       \end{cases}\label{eq:phibelow}
\end{align}
where the evaluation of the limit was carried out with the same
technique based on Lemma~\ref{thm:ptmeas} used in the previous
sections. Figure~\ref{fig:phibound} displays the lower bounds
(\ref{eq:phicrude}) and (\ref{eq:phibelow}) together with the
exact result (\ref{eq:20}).

It is apparent that these bounds are rather weak, and in fact
completely trivial for large rates. Hence all-particle fidelities
contain new and independent information about purification
processes, which is not already contained in their one-particle
counterparts.

\section*{Acknowledgements}

We acknowledge several rounds of email discussions with Ignacio
Cirac, which helped to clarify the precise relation between this
work and \cite{IAC}. 


 \end{document}